# PERPENDICULARLY BIASED YIG TUNERS FOR THE FERMILAB RECYCLER 52.809 MHZ RF CAVITIES*

R. Madrak#, V. Kashikhin, A. Makarov, D. Wildman, FNAL, Batavia, IL 60510, USA

*Abstract*

For NOvA and future experiments requiring high intensity proton beams, Fermilab is in the process of upgrading the existing accelerator complex for increased proton production. One such improvement is to reduce the Main Injector cycle time, by performing slip stacking, previously done in the Main Injector, in the now repurposed Recycler Ring. Recycler slip stacking requires new tuneable RF cavities, discussed separately in these proceedings. These are quarter wave cavities resonant at 52.809 MHz with a 10 kHz tuning range. The 10 kHz range is achieved by use of a tuner which has an electrical length of approximately one half wavelength at 52.809 MHz. The tuner is constructed from 3⅛″ diameter rigid coaxial line, with 5 inches of its length containing perpendicularly biased, Al doped Yttrium Iron Garnet (YIG). The tuner design, measurements, and high power test results are presented.

## INTRODUCTION

Two new quarter wave copper cavities have been constructed and installed in the Fermilab Recycler Ring for the purpose of slip stacking protons to increase beam power. These have been described in detail in [1]. They are to operate at 52.809 MHz with a tuning range of 10 kHz. Tuning by 1260 Hz is required for slip stacking itself, and additional tuning is needed to compensate for temperature changes. The cavity maximum peak gap voltage is 150 kV with a shunt impedance of 75 kΩ and Q ~ 6000. The cavities will be pulsed for a maximum of 0.8 s every 1.33 s (the Main Injector cycle time).

The cavities are tuned using perpendicularly biased garnet in a coaxial line. This has been done previously and is described in [2] and [3]. The original concept, though not for cavity tuning in particular, is described in [4].

## DESIGN

In one of the ports located 3″ (on center) from the cavity shorted end, a fast tuner is loop-coupled to the cavity. The coupling is through a 2.5″ diameter coaxial ceramic window. The tuner loop area of 2.5 in$^2$ was adjusted to give a nominal coupling impedance of 50 Ω. The tuner is slightly less than a half-wavelength long and is made up of standard EIA 3⅛″ diameter, 50 Ω rigid copper transmission line, in addition to an "adjustable section" which is partially loaded with a 5″ long piece of Al doped Yttrium Iron Garnet (YIG). This section is immersed in a variable solenoidal magnetic field which provides a perpendicular bias for cavity tuning. A photo of the adjustable section, with the center conductor assembly removed from the outer conductor, is shown in Figure 1. The tuner line is shorted at the end opposite the cavity.

The garnet (TCI Ceramics type AL-400) has a saturation magnetization ($4\pi M_s$) of 400 gauss. Its OD and ID are 3.0″ and 0.65″ respectively. The center conductor used in the 15.75″ long adjustable section, which does not have the dimensions of standard 3⅛″ coax line, is shrink fit into the garnet. Over the 9″ length closest to the short, the outer conductor thickness has been reduced from the standard 3⅛″ coax line wall thickness to 0.020″. Both this 9″ section and the bottom copper shorting plate have a 0.0197″ wide slot machined through the copper to reduce eddy current effects.

For heat removal, the outer conductor is water cooled and the adjustable section is filled with a dielectric fluid (Diala AX).

The adjustable section had previously been designed for use as a fast phase shifter in vector modulators operating at 325 MHz [5].

The entire tuner, attached to the cavity and installed in the Recycler, is shown in Figure 2.

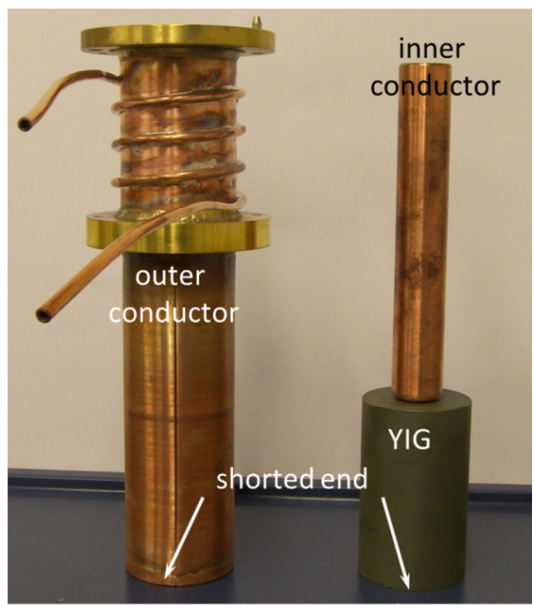

Figure 1: Tuner adjustable section with the inner conductor removed from the outer conductor.

___________________

*Solenoid*

The 112 turn solenoid (with TCI Ceramics G4 ferrite flux return) was designed to work at relatively high DC current (80A), but also at high frequency (~2 kHz) AC current. While the AC response is not absolutely required, the ability to tune at ~2 kHz will be an advantage. Obtaining the desired DC and AC response necessitated the use of Litz Cable of sufficient cross section (1/0 AWG) with 270 copper strands of 24 AWG wire insulated from each other. The total strand cross section reduces the losses in the DC regime, and the insulation between strands minimizes the eddy currents in the cable (for the AC regime). A cooling tube was wound on the outer surface of the solenoid as a bifilar coil in order to eliminate the effect of magnetic fields caused by eddy currents induced in it. A DC magnetic field simulation predicted the resulting minimum and maximum magnetic fields in the tuner garnet of 0.058 T and 0.081 T, for 8000 solenoid ampere-turns.

The solenoid current is supplied by a Copley Controls Model 266 amplifier which has a nominal bandwidth of

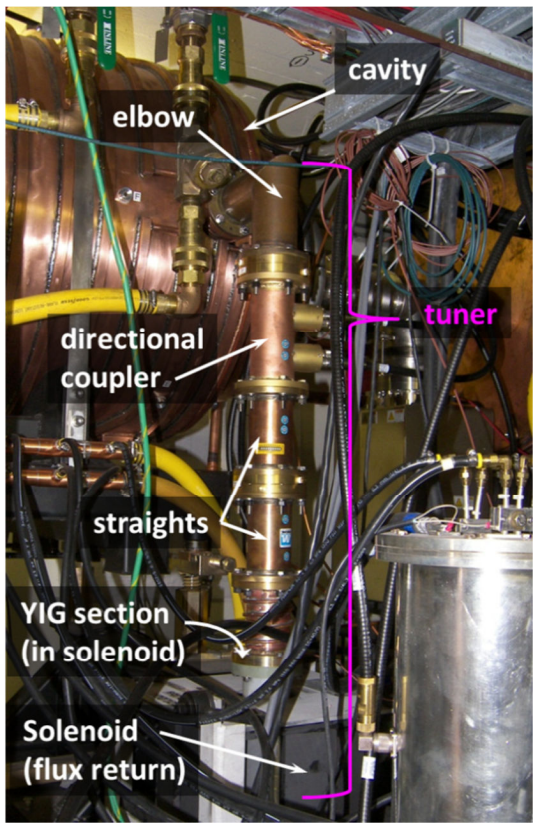

Figure 2: Tuner attached to the cavity, installed in the Recycler. The tuner is made up several 3⅛″ coax sections: an elbow, directional coupler, several straight sections, and finally the YIG loaded adjustable section (here obscured by the solenoid) which is shown in Figure 1.

5 kHz. The tuner response is currently limited by the solenoid inductance of 1.9 mH and the 60V max/80 A max DC supply for the Copley amplifier.

# TUNER PERFORMANCE AND MEASUREMENTS

The tuner is characterized apart from the cavity by the measurements plotted in Figure 3. This shows network analyzer $S_{11}$ amplitude and phase measurements as a function of DC solenoid bias. Large phase shifts correspond to a large frequency change (tuning) in the cavity/tuner system. Operation must be above ~35A, below which the garnet becomes lossy.

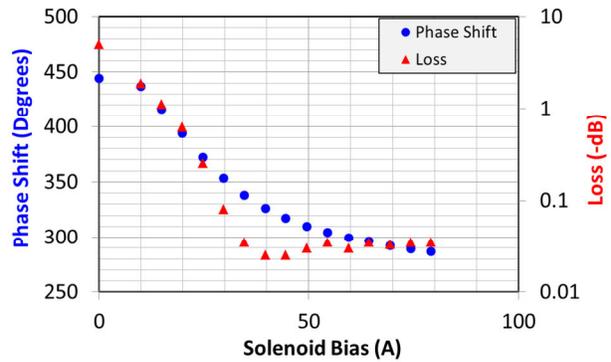

Figure 3: Two-way phase shift in tuner 15.75″ adjustable section at 52.809 MHz (left y-axis) and Loss (right y-axis) as a function of solenoid bias.

The next set of measurements quantifies the operation of the cavity/tuner system as a whole. Figure 4 (wide range) and Figure 5 (operation range) show the frequency (left y-axis) and the Q of the cavity/tuner system (right y-axis) as a function of solenoid bias current. The bias at which we observe a large change in frequency and a minimum in Q corresponds to the case where the tuner is one half wavelength long.

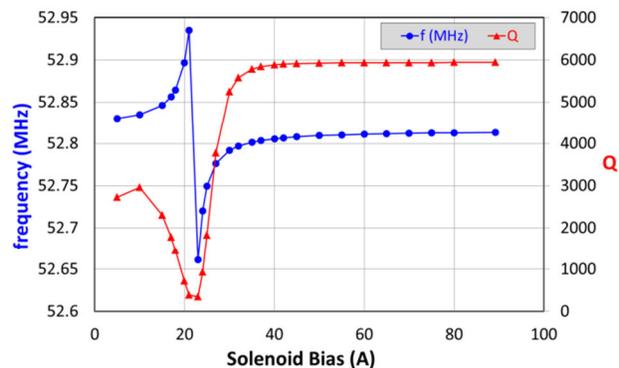

Figure 4: Frequency and Q of the cavity/tuner system as a function of solenoid bias current, for a wide range of tuner biases.

An additional restriction on tuner range is due to the peak voltage that can be sustained in the tuner without sparking. Figure 6 shows low level measurements of frequency and the corresponding tuner voltage. In this case, the cavity was powered with a relatively small signal. Cavity peak accelerating gap voltage and tuner peak voltage were measured simultaneously. The values were then scaled to cavity gap voltages of 100 and 150 kV. Operating with a tuner voltage above the peak voltage limit for 3⅛″ coaxial line of 13.3 kV will cause arcing and must be avoided. Not only is arcing in the coax line undesirable, but this can also lead to high voltages near the garnet which could be destructive.

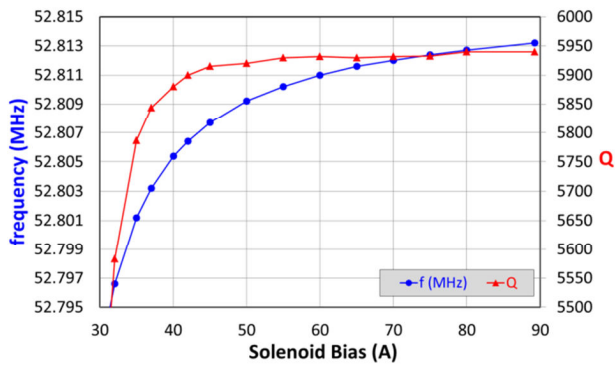

Figure 5: Frequency and Q of the cavity/tuner system as a function of solenoid bias current, for solenoid biases in the range that the tuner is operated (> 35 A).

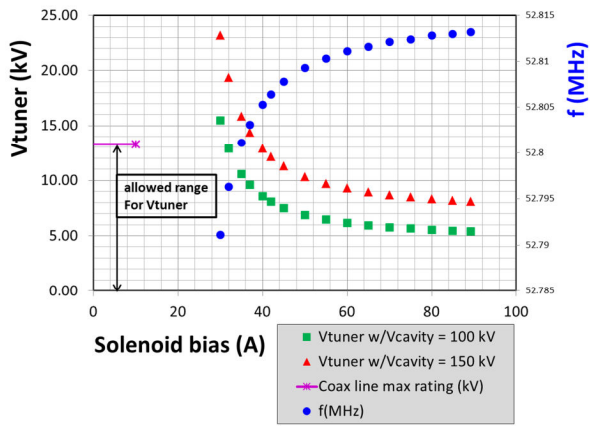

Figure 6: Tuner voltage and frequency of cavity/tuner system as a function of tuner bias. The maximum peak voltage for the tuner 3⅛″ line is also shown. (These measurements were taken from the cavity field probe and the directional coupler which is part of the tuner.)

Measurements shown indicate I ≥ 32 A and I ≥ 40 A to respect the 13.3 kV limit, for cavity voltages of 100 kV and 150 kV. With no other adjustments, and taking into account the 35 A limit imposed due to losses in the garnet, the tuning ranges are 11.7 kHz and 7.5 kHz for cavity peak voltages of 100 kV and 150 kV.

## TUNER SPARK PROTECTION

To avoid destruction of the garnet, several methods of protection are in place to shut off RF power to the cavity in the event of a tuner spark.

The tuner line contains a 60 dB directional coupler with both forward and reverse power taps. The first tuner protection system shuts off the RF drive if a change of more than 3dB/μs is detected in the reverse power.

In the second protection system, a phase detector monitors the phase difference between forward and reverse power. If the phase changes more quickly than that due to normal tuning, the RF drive is again shut off. This is accomplished using a capacitive differentiator and a comparator.

## STATUS AND HIGH POWER TESTING

Before being installed in the Recycler, both cavities were high power tested in a shielded "cavity test cave". The cavity and tuner systems have been operated successfully at high power in the Recycler.

## ACKNOWLEDGMENTS

We would like to acknowledge the huge efforts of the Accelerator Division Mechanical Support and RF Groups, and everyone involved in this project. In addition, we would like to thank all of those in Technical Division involved in the design and construction of the solenoids.